\documentclass{llncs}
\usepackage[utf8]{inputenc}
\usepackage{amsmath,amssymb,latexsym}
\usepackage{mathtools}
\usepackage{xspace}
\usepackage{color}
\usepackage[colorlinks]{hyperref}
\usepackage{float}
\floatstyle{boxed}
\restylefloat{figure}
\usepackage{todonotes}
\usepackage{caption}
\usepackage{subcaption}

\usepackage{bussproofs}
\EnableBpAbbreviations

\newcommand{\UICm}[1]{\UIC{$#1$}}
\newcommand{\AXCm}[1]{\AXC{$#1$}}

\newcommand{\RLm}[1]{\RL{$#1$}}

\newcommand{\LLm}[1]{\LL{$#1$}}

\newcommand{\myfig}[1]{Figure~#1}

\def\rrightarrow{\twoheadrightarrow}

\def\brsep{\; |\;}

\def\calR{{\cal R}}

\def\gs3{{\sf GS3}\xspace} 

\def\limp{\Rightarrow}
\def\land{\wedge}
\def\lor{\vee}
\def\lfa{\forall}

\def\lex{\exists}

\title{Polarized Rewriting and Tableaux in B Set Theory}
\author{Olivier Hermant}
\institute{MINES ParisTech, PSL University, France \\
 \email{olivier.hermant@mines-paristech.fr}}

\DeclareMathAlphabet{\mathpzc}{OT1}{pzc}{m}{it}
\begin{document}
\maketitle

\begin{abstract}
We propose and extension of the tableau-based first-order
automated theorem prover Zenon Modulo to polarized rewriting. We
introduce the framework and explain the potential benefits. The first
target is an industrial benchmark composed of B Set Theory problems.
\end{abstract}

\section{Introduction}

The B Method set theory \cite{JRAbr96} has been extensively used for
20 years by the railway industry in France to develop certified
correct-by-construction software. Recently, the BWare \cite{BWar12}
project has tackled the issue of automatically proving the
thousands of \emph{proof obligations} generated by the development
process.

Zenon Modulo \cite{DDelDDolFGilPHalOHer13} is one of the tools
developed to this aim. Originally a tableau-based prover, Zenon
\cite{RBonDDelDDol07} is used for instance by TLA+
\cite{DCouDDolLLamSMerDRicHVan12} and FoCaLiZe \cite{CDubRRio14}. To
help manage the axioms of set theory, but also the uncountable derived
constructs definitions (e.g. inclusion, union, functions), we deemed
useful to not let nonlogical axioms wander as formulas: a prover would
easily get lost by decomposing one or another axiom in an unorganized
fashion.

We replaced them with rewrite rules, turning Zenon into an
implementation of Deduction Modulo Theory
\cite{RBon04,RBonOHer06a,GDowTHarCKir03}, which allows rewriting on
terms and \emph{formulas}. Additionally, we equipped it with ML
polymorphic types and arithmetic \cite{GBurDDelDDolPHalOHer15}. On the
BWare benchmark, the success rate was raised from 2.5\% to 95\%.

We propose to extend Zenon Modulo with \emph{polarized} rewriting, a
more permissive rewrite relation. We first introduce the framework,
then we discuss examples and the pros and cons of the approach. There
is currently no implementation, essentially because this is a perfect
match for an intern or a PhD student.

\section{Polarized Tableaux Modulo Theory}

We assume familiarity with first-order logic and at
least one deduction system. Tableaux calculus is a refutational
calculus, thus, to show $F$ under the assumptions $\Gamma$, we
refute $\Gamma, \neg F$.  The first-order tableaux rules are recalled
in \myfig{\ref{fig:tab}}, see textbooks \cite{ANerRsho93} for details.
The rules have the following characteristics:
\begin{itemize}
\item as customary, they are presented in a top-down
  fashion.
\item Formulas are not in negation normal form, rules are duplicated.
\item A branch may be closed, denoted $\odot$, if we find on it
  (including internal nodes) an occurrence of some $F$ and its
  negation, or an explicit contradiction. A tableau is a proof iff
  each branch is closed.
\item $\alpha$-rules are for non-branching connectives rules and
  $\beta$-rules for branching ones, $\delta$-rules are for quantifier
  rules introducing a fresh constant $c$ and $\gamma$-rules for those
  introducing any term.
\end{itemize}

\begin{figure}[h]
\parbox{\textwidth}{
\begin{center}
  \begin{tabular}{c@{\hspace{0.4cm}}c@{\hspace{0.4cm}}c@{\hspace{0.4cm}}c}
    \multicolumn{4}{c}{
      \begin{tabular}{c@{\hspace{1.2cm}}c@{\hspace{1.2cm}}c}
        \rootAtTop
        \AXCm{\odot}\RLm{\odot_\bot}
        \UICm{\bot}\DP &
        \rootAtTop
        \AXCm{\odot}\RLm{\odot}
        \UICm{F,\neg{}F}\DP
        & 
        \rootAtTop
        \AXCm{\odot}\RLm{\odot_{\neg\top}}
        \UICm{\neg\top}\DP
      \end{tabular}
    }
    \\\\
    \rootAtTop
    \AXCm{F}\RLm{\alpha_{\neg\neg}}
    \UICm{\neg\neg{}F}\DP
    &
    \rootAtTop
    \AXCm{F,G}\RLm{\alpha_\land}
    \UICm{F\land{}G}\DP
    &
    \rootAtTop
    \AXCm{\neg{}F,\neg{}G}\RLm{\alpha_{\neg\lor}}
    \UICm{\neg(F\lor{}G)}\DP
    &
    \rootAtTop
    \AXCm{F,\neg{}G}\RLm{\alpha_{\neg\Rightarrow}}
    \UICm{\neg(F\Rightarrow{}G)}\DP
    \\\\
    \multicolumn{4}{c}{
      \begin{tabular}{c@{\hspace{1cm}}c@{\hspace{1cm}}c}
        \rootAtTop
        \AXCm{F\brsep{}G}\RLm{\beta_\lor}
        \UICm{F\lor{}G}\DP
        &
        \rootAtTop
        \AXCm{\neg{}F\brsep\neg{}G}\RLm{\beta_{\neg\land}}
        \UICm{\neg(F\land{}G)}\DP
        &
        \rootAtTop
        \AXCm{\neg{}F\brsep{}G}\RLm{\beta_\Rightarrow}
        \UICm{F\Rightarrow{}G}\DP
      \end{tabular}
    } \\\\
    \multicolumn{4}{c}{
      \begin{tabular}{c@{\hspace{2cm}}c}
        \rootAtTop
        \AXCm{F(c)}\RLm{\delta_\exists}
        \UICm{\exists{x}\;F(x)}\DP
        &
        \rootAtTop
        \AXCm{\neg{}F(c)}\RLm{\delta_{\neg\forall}}
        \UICm{\neg\forall{x}\;F(x)}\DP
        \\\\
        \rootAtTop
        \AXCm{F(t)}
        \RLm{\gamma_{\forall}}
        \UICm{\forall{x}\;F(x)}\DP
        &
        \rootAtTop
        \AXCm{\neg{}F(t)}
        \RLm{\gamma_{\neg\exists}}
        \UICm{\neg\exists{x}\;F(x)}\DP
      \end{tabular}
    }
  \end{tabular}
\end{center}
}
\caption{Tableaux Rules}
\label{fig:tab}
\end{figure}

Tableaux Modulo Theory \cite{RBon04} extends tableaux with a set of
rewrite rules $\calR$. A rewrite rule is a pair of terms, $l
\rightarrow r$, where the variables of $r$ appear in $l$. Given a set
$\calR$, a term $t$ rewrites into $u$, denoted $t \rightarrow
u$, if there is a rule $l \rightarrow r \in \calR$ and a substitution
$\sigma$, such that there is an occurrence of $l\sigma$ in $t$, and
$u$ is $t$ where that occurrence has been replaced with $r\sigma$. In
other words, $\rightarrow$ is the closure of $\calR$ by substitution
and the subterm relation. The transitive closure of $\rightarrow$ is
denoted $\rrightarrow$ and its further reflexive-symmetric closure is
$\equiv$, which is a congruence.

Deduction Modulo Theory also allows rewrite rules on \emph{formulas},
provided the left member $P$ of such a rule $P \rightarrow F$ is
atomic. The relations $\rightarrow, \rrightarrow, \equiv$ on formulas
embed their counterparts on the subterms of the formulas.

Tableaux can be extended to rewriting with the addition of a rule
allowing to convert any formula with $\equiv$, as in
\myfig{\ref{fig:conv}}. When rewriting is confluent, we can orient
this rule as in \myfig{\ref{fig:confluent-conv}}. In practice,
Deduction Modulo Theory-based automated theorem provers \cite{GBur10}
implement this last rule, which is a way to decide $\equiv$ when
confluence holds. Other presentations exist
\cite{RBon04,RBonOHer06a,GBurDDelDDolPHalOHer15}.

\begin{figure}[htb!]
  \parbox{\textwidth}{
    \begin{center}
    \begin{subfigure}[]{0.4\textwidth}
    \begin{prooftree}
      \AXCm{F}
      \RLm{\equiv\mbox{, if } F \equiv G}
      \UICm{G}
    \end{prooftree}
    \caption{General Case}
    \label{fig:conv}
    \end{subfigure}
    ~
    \begin{subfigure}[]{0.4\textwidth}
    \begin{prooftree}
      \AXCm{F}
      \RLm{\rrightarrow\mbox{, if } F \rrightarrow G}
      \UICm{G}
    \end{prooftree}
    \caption{Confluent Case}
    \label{fig:confluent-conv}
    \end{subfigure}
    \end{center}
  }
  \caption{The Additional Rule of Tableaux Modulo Theory}
  \label{fig:conv-all}
\end{figure}

The calculus of Zenon Modulo \cite{GBurDDelDDolPHalOHer15} enjoys
meta-variables, Hilbert's $\epsilon$ operator, reasoning over
reflexive/transitive/symmetric relations, an equality predicate,
ML-polymorphic types, and, of course, rewriting. The simpler
case of \myfig{\ref{fig:tab}} is sufficient here, as we focus
on rewriting, that we now extend to polarity.

\begin{definition}[Polarity of an Occurrence]
  The occurrence of a formula $F$ in a formula $G$ is positive
  (resp. negative) iff
  \begin{itemize}
  \item $G$ is $F$,
  \item $G$ is $G_1 \land G_2$, $G_1 \lor G_2$, $\lfa{x} G_1$,
    $\lfa{x} G_1$ or $H \limp G_1$ and the occurence of $F$ in
    $G_1$ or $G_2$ is positive (resp. negative),
  \item $G$ is $\neg G_1$ or $G_1 \limp H$ and the occurrence of $F$ in $G_1$ is
    negative (resp. positive).
  \end{itemize}
\end{definition}

\noindent Now, we consider two (proposition) rewrite systems $\calR^+ \cup
\calR^-$.

\begin{definition}[Polarized Rewrite Relation]
  Let $F$ and $G$ be two formulas. $F \rightarrow_+ G$ iff $F
  \rightarrow G$ with a term rewrite rule or there exists a positive
  (resp. negative) occurrence $H$ in $F$, a substitution $\sigma$, and
  a rule $l \rightarrow r \in \calR^+$ (resp. $\calR^-$), such that $H
  = l\sigma$ and $G$ is $F$ where $H$ has been replaced with
  $r\sigma$.

  $F \rightarrow_- G$ iff $\neg F \rightarrow_+ \neg G$, that is to
  say we exchange $\calR^+$ and $\calR^-$ above.
\end{definition}

  We denote by $\rrightarrow_+$ and $\rrightarrow_-$ the
  reflexive-transitive closures of $\rightarrow_+$ and
  $\rightarrow_-$, respectively. Defining $\equiv_+$ and $\equiv_-$ is
  more delicate and unlikely to be useful practice. Polarized Tableaux
  combine the rules of \myfig{\ref{fig:tab}} and
  \myfig{\ref{fig:polarized-conv}}.

\begin{figure}[ht!]
  \parbox{\textwidth}{
    \begin{prooftree}
      \AXCm{F}
      \RLm{\rrightarrow_+\mbox{, if } F \rrightarrow_+ G}
      \UICm{G}
    \end{prooftree}
  }
  \caption{The Additional Rule of Polarized Tableaux Modulo Theory}
  \label{fig:polarized-conv}
\end{figure}

\section{Implementation}

Zenon Modulo rewrites only literals, in a forward fashion. This is a
further restriction of \myfig{\ref{fig:confluent-conv}} and it relies
on termination of term rewriting and on confluence of the whole
rewriting. Otherwise, completeness of the proof search fails. The
heuristic is, each time we meet a literal, to:
\begin{enumerate}
\item normalize the terms it contains;
\item rewrite the literal itself (if there is an applicable rewrite
  rule) on \emph{one step};
\item if the formula is in normal form or compound, stop, otherwise
  repeat.
\end{enumerate}

To get polarized rewriting it suffices to modify the second step into
``rewrite positively if the literal is positive, negatively
otherwise''. The expected gain does not lie here, but in an
\emph{optimized preprocessing} for rules.  Indeed \cite{GDow10}, a
polarized rule $P \rightarrow_+ F \in \calR^+$ represents/can
be represented as an axiom $\overline{\lfa} (P \limp F)$
($\overline{\lfa}$ is the universal closure over the free
variables). Similarly, a negative rewrite rule $P \rrightarrow_- F \in
\calR^-$ is equivalent to the axiom $\overline{\lfa} (F \limp P)$. In
contrast, Deduction Modulo Theory's rewrite rules $P \rightarrow F$
are equivalent to $\overline{\lfa} ((P \Leftrightarrow F)$
\cite{GDowTHarCKir03}. Remind that we are discussing propositional
rewrite rules, so $P$ has to be atomic. Consequently, polarization
offers the following improvements:
\begin{itemize}
\item more axioms correspond to rewrite rules, and this improves proof
  search \cite{DDelDDolFGilPHalOHer13}. Axioms of the form $\lfa
  \overline{x} (P \limp A)$ and $\lfa \overline{x} (A \limp P)$, with
  $P$ atomic, become rules of $\calR^+$ and $\calR^-$,
  respectively. In classical logic, when $P$ is a negated atom, we
  also get rewrite rules in $\calR^-$ and $\calR^+$, respectively.
  
\item We can \emph{Skolemize} rewrite rules. This has two
  benefits: first, less inference rules are necessary in the
  tableaux, and second, the Skolem term is \emph{uniform}, while
  multiple applications of $\delta_\lex$ or $\delta_{\neg\lfa}$
  introduce different fresh symbols at each time. This also
  holds in the presence of meta-variables.
\end{itemize}

Skolemizing the rules is impossible in vanilla Deduction Modulo
Theory, as rewriting applies at positive and negative
occurrences. Therefore, we do not know in advance which quantifiers
are positive and negative. To illustrate the difference, consider
axioms of the type $\lfa \overline{x} (P \limp A)$ and $\lfa
\overline{x} (A \limp P)$.
\begin{itemize}
  \item In $\lfa  \overline{x} (P \limp A)$, we can replace
    all the positive existential and negative
    universal quantifiers of $A$ by a Skolem function symbol.
  \item Similarly, in $\lfa \overline{x} (A \limp P)$, we can
    replace all the positive universal and
    negative existential quantifiers of $A$ by a Skolem function
    symbol.
\end{itemize}

The very same principle applies to polarized rewrite rules. We leave
the study and the choice of the strategies for Skolemization
\cite{ANonCWei01} for a later stage.  Both improvements can be applied
to heuristics turning assumptions (of a given problem) into rewrite
rules, and to hand-tuning of the rewrite rules of a specific theory,
for instance B Method set theory.

\section{Example}

\begin{figure}[htb!]
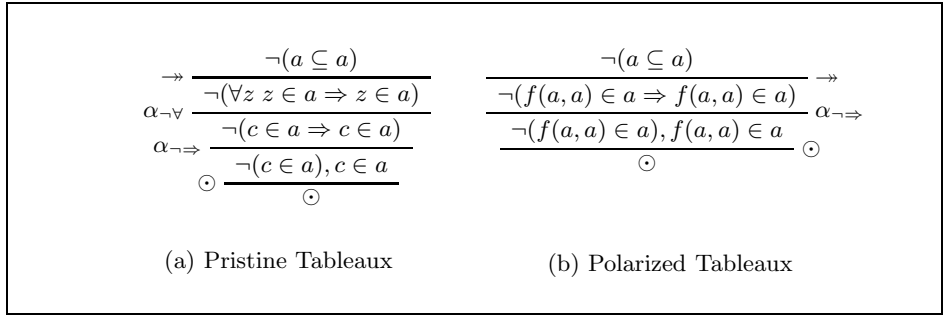

  \parbox{\textwidth}{
    \begin{center}
    \begin{subfigure}[]{0.4\textwidth}
      \begin{prooftree}
        \rootAtTop
        \AXCm{\odot}
        \LLm{\odot}
        \UICm{\neg (c \in a), c \in a}
        \LLm{\alpha_{\neg\limp}}
        \UICm{\neg(c \in a \limp c \in a)}
        \LLm{\alpha_{\neg\lfa}}
        \UICm{\neg(\lfa{z}\; z \in a \limp z \in a)}
        \LLm{\rrightarrow}
        \UICm{\neg (a \subseteq a)}
      \end{prooftree}
    \caption{Pristine Tableaux}
    \label{fig:tmt}
    \end{subfigure}
    ~
    \begin{subfigure}[]{0.4\textwidth}
      \begin{prooftree}
        \rootAtTop
        \AXCm{\phantom{\neg (a \subseteq a)}}
        \noLine
        \UICm{\odot}
        \RLm{\odot}
        \UICm{\neg (f(a,a) \in a), f(a,a) \in a}
        \RLm{\alpha_{\neg\limp}}
        \UICm{\neg(f(a,a) \in a \limp f(a,a) \in a)}
        \RLm{\rrightarrow}
        \UICm{\neg (a \subseteq a)}
      \end{prooftree}
      \caption{Polarized Tableaux}
    \label{fig:ptmt}
    \end{subfigure}
    \end{center}
  }
  \caption{Proof of $a \subseteq a$ in Deduction Modulo Theory}
  \label{fig:ainca}
\end{figure}

Let us consider the classical example of proving $a \subseteq a$ with
the standard axiom of inclusion $\forall{x} \forall{y}\;x \subseteq y
\Leftrightarrow (\lfa{z}\;z \in x \limp z \in y)$. A usual tableau
proof involves the succession of rules $\gamma_\lfa$ (twice),
$\alpha_\land$, $\beta_\limp$, $\delta_{\neg\lfa}$ and
$\alpha_{\neg\limp}$ on the axiom. Deduction Modulo Theory turns it
into the rewrite rule $ x \subseteq y \rightarrow (\lfa{z}\;z \in x
\limp z \in y)$, and yields the 3-rules axiomless proof of
\myfig{\ref{fig:tmt}}.

If we switch to Polarized Deduction Modulo Theory, we get the pair
$\calR^+ = \{ x \subseteq y \rightarrow (\lfa z z \in x \limp z \in y)
\}$ and $\calR^- = \{ x \subseteq y \rightarrow (f(x,y) \in x \limp
f(x,y) \in y)\}$. The proof of $a \subseteq a$ is one more
step smaller, as shown in \myfig{\ref{fig:ptmt}}.

\section{Conclusion}

We expect the polarized approach to give at least as efficient as
Zenon Modulo itself. The proof-search algorithm needs only few
changes, mostly in the rule
pre-processing. The obtained rules contain less quantifiers,
allowing for fewer rules in proof-search and potential
earlier unification and branch closure, since using a rewrite rule
several times now involves the \emph{same} Skolem symbol.

On the risk side, implicational axioms can now be turned into rewrite
rules. This might be a threat to termination or confluence. A study of
the theoretical framework, including models, is required.\\

Automated theorem provers are aggressively optimized tools, naturally
lending themselves to bugs. This is why independent double checking
facilities are important. Zenon Modulo is able to produce
\emph{proof-terms} or \emph{proof certificates}, though it provides no
rewrite steps explicitly, following Poincaré's Principle: computations
(rewriting) in proofs give no insight, they can be quickly
reconstructed (by the checker) at will and are to be left
implicit. Such a clerk/expert distinction has for instance been
studied in the Foundational Proof Certificate project \cite{DMil15},
at the proof level, with the help of focusing \cite{ZChiDMilFRen17}.

On the BWare benchmark, all statements proved by Zenon Modulo
\cite{GBurDDelDDolPHalOHer15}, that do not involve arithmetic, are
actually declared well-typed by Dedukti \cite{MBoeQCarOHer12}, a type
checker based on an extension of Deduction Modulo Theory to dependent
types. Dedukti's rewriting ability made extremely smooth the
reconstruction of rewriting : there is essentially nothing to do but
to declare the rules.

The challenge is to keep this skeptical double-checking approach
viable. We may need a \emph{depolarization} of the proofs, perhaps
following \cite{GBurCKir10}, or an substantial extension of Dedukti
and its foundations to polarized rewriting, perhaps with the help of
subtyping.



\bibliographystyle{splncs03}
\bibliography{biblio}
\end{document}